\documentclass[conference]{IEEEtran}
\IEEEoverridecommandlockouts
\usepackage{cite}
\usepackage{amsmath,amssymb,amsfonts}
\usepackage{algorithm}
\usepackage{algpseudocode}
\usepackage{graphicx}
\graphicspath{ {./figures/} }
\usepackage[caption=false, font=footnotesize]{subfig}
\usepackage{textcomp}
\usepackage{xcolor}
\def\BibTeX{{\rm B\kern-.05em{\sc i\kern-.025em b}\kern-.08em
    T\kern-.1667em\lower.7ex\hbox{E}\kern-.125emX}}

\usepackage{fancyhdr}

\begin{document}

\title{Optimal Placement of Limited PMUs for Transmission Line Outage Detection and Identification\\
\thanks{This work is partially supported by the Future Resilient Systems Project at the Singapore-ETH Centre (SEC), which is funded by the National Research Foundation of Singapore (NRF) under its Campus for Research Excellence and Technological Enterprise (CREATE) program. 
\copyright 2020 IEEE. Personal use of this material is permitted. Permission from IEEE must be
obtained for all other uses, in any current or future media, including
reprinting/republishing this material for advertising or promotional purposes, creating new
collective works, for resale or redistribution to servers or lists, or reuse of any copyrighted
component of this work in other works.}
}

\author{\IEEEauthorblockN{Xiaozhou Yang and Nan Chen}
\IEEEauthorblockA{Department of Industrial Systems
Engineering and Management \\
National University of Singapore \\
Singapore, Singapore \\
xiaozhou.yang@u.nus.edu}
\and
\IEEEauthorblockN{Chao Zhai}
\IEEEauthorblockA{School of Automation \\
China University of Geosciences \\
Wuhan, China \\
zhaichao@amss.ac.cn}}

\maketitle

\thispagestyle{fancy}
\fancyhead{}
\lhead{}
\lfoot{978-1-7281-2822-1/20/\$31.00~\copyright2020 IEEE \hfill}
\cfoot{}
\rfoot{PMAPS 2020}
\renewcommand{\headrulewidth}{0pt}
\renewcommand{\footrulewidth}{0pt}

\begin{abstract}
Phasor Measurement Unit (PMU) technology is increasingly used for real-time monitoring applications, especially line outage detection and identification (D\&I) in the power system. Current outage D\&I schemes either assume a full PMU deployment or a partial deployment with fixed PMU placement. However, the placement of the PMUs has a fundamental impact on the effectiveness of the D\&I scheme. Building on a dynamic relationship between the substation voltage phase angle and active power, we formulated the optimal PMU placement problem for outage D\&I as an optimization problem readily solvable by any heuristic algorithm. We tested the formulation using a genetic algorithm and simulated outages of IEEE 39 bus system. The optimal placement found produces a better D\&I result of single-line outages than a randomly scattered, tree-like, and degree-based placements. 
\end{abstract}


\section{Introduction}
Power systems are critical infrastructures essential to modern livelihood. It is incredibly complex because of the extensive geographical scale, fast dynamics, and high operational standards. There is also increasing volatility in power systems due to the integration of distributed energy resources. Independent system operators (ISOs) demand more intelligent real-time monitoring tools to detect and locate abnormal events and minimize the economic impact of such events. One common and extensively researched abnormal event in power systems is transmission line outage. Line outages can happen due to various reasons, such as severe weather conditions, equipment failures, and component wear and tear. 

Outage dynamics propagate through systems in a time scale of milliseconds, and traditional supervisory control and data acquisition devices are not able to capture these dynamics \cite{pignati2015real}. Recognizing its real-time monitoring capability, ISOs are progressively installing phasor measurement units (PMUs) on their power grids.
PMUs are devices installed at substations, capable of recording high-precision, high-fidelity, and GPS time-synchronized measurements. An industry-grade PMU could measure substation current and voltage phasors with a total vector error of less than $1\%$ according to the IEEE C37.118.1-2011 standard. The sampling frequency could also reach between 30 to 60 samples per second. Since its introduction, researchers have been studying PMU technology for tasks such as dynamic state estimation, stability control, and fault detection. See \cite{Aminifar2014} for a comprehensive review of PMU applications in power systems. 

Recent literature focuses on using PMU data for real-time line outage detection and identification (D\&I) in power systems \cite{Tate2008, Rafferty2016, Chen2016, Hosur2019, Ardakanian2019a, yang2019control}. However, they either assume PMUs are installed on all the buses, or a limited number of PMUs are installed on pre-determined locations. Due to the economic and data-handling constraint, ISOs need to work with a limited number of PMUs, i.e., some parts of the system are unobservable. The placement of PMUs can influence the effectiveness of the outage D\&I scheme. Therefore, it is necessary to investigate the optimal PMU placement (OPP) problem for the specific applications.

OPP problems are combinatorial optimization problems since there are $2^N$ possible placements for a system with N buses. Traditionally, many researchers focus on finding placements optimal for network observability, using variants of heuristic algorithms, e.g., genetic algorithm \cite{marin2003genetic}, simulated annealing \cite{dua2008optimal}, and Tabu search \cite{koutsoukis2013numerical}. In a work close to ours \cite{Geramian2008}, the authors formulated a placement optimization problem so that the minimal number of PMUs are installed to ensure complete fault observability. However, they focus on the location of the fault on the transmission line, whereas our work focuses on line outage D\&I. Recently, OPP problems are studied for dynamic state estimation \cite{qi2014optimal}, bad data detection \cite{gou2014unified}, as well as anomaly detection and localization \cite{rajeev2015fault, Jamei2017a}. In this work, we formulate the OPP problem as an optimization problem where the placement found ensures a minimal approximation error between the system model and the actual system behavior. This formulation finds explicitly a placement such that the accuracy of line outage D\&I can be improved given a limited number of PMUs. 

The rest of this paper is organized as follows. Section \ref{sec:methodology} introduces the formulation of a dyanmic power system model and the genetic algorithm for D\&I scheme performance improvement. Then, simulation results are shown in Section \ref{sec:simulation} to demostrate the effectiveness of the proposed method. We summarize the findings in Section \ref{sec:conclusion} and suggest some future research directions.

\section{Methodology}
\label{sec:methodology}
\subsection{System Model with Limited PMUs}
Given a power system with $N$ buses connected by $L$ transmission lines, the power grid network can be modeled as a graph $\mathcal{G} = (\mathcal{N}, \mathcal{E})$, $\mathcal{N} = \{1, 2, \cdots, N\}$, and $\mathcal{E} \subseteq N  \times N$ where $\mathcal{N}$ is the set of N buses and $\mathcal{E}$ is the set of L transmission lines. For every bus, let P be the net active power, Q be the net reactive power, V be the nodal voltage magnitude, and $\theta$ be the phase angle. Power flows in the network can be described by the alternate current (AC) power flow model:
\begin{subequations}
\label{eqn:AC_power_flow_model}
\begin{align}
\text{P}_m &= \text{V}_m \sum_{n=1}^{N} \text{V}_n \text{Y}_{mn} \cos (\theta_m - \theta_n - \alpha_{mn}) \,, \label{eqn:AC_power_flow_P}\\
\text{Q}_m &= \text{V}_m \sum_{n=1}^{N} \text{V}_n \text{Y}_{mn} \sin (\theta_m - \theta_n - \alpha_{mn}) \,, \label{eqn:AC_power_flow_Q}
\end{align}
\end{subequations}
for bus $m = 1, 2, \dots, N$ \cite{Glover2012}. Y$_{mn}$ is the magnitude of the $(m,n)_{th}$ complex admittance of the bus admittance matrix when it is written in the exponential form. Linearizing and retaining the real power portion of Eqn. \ref{eqn:AC_power_flow_model} in the same way as \cite{yang2019control}, we obtain a discrete-time dynamic relationship between \textbf{P} and $\boldsymbol{\theta}$, two $(N-1)$-vectors by removing the reference bus, as:
\begin{equation}
\label{eqn:small_signal_model}
    \Delta \textbf{P}_k =  \mathbf{J}(\boldsymbol{\theta}_{k-1}) \Delta \boldsymbol{\theta}_k \,,
\end{equation}
where $\Delta \textbf{P}_k = \textbf{P}_k - \textbf{P}_{k-1}$ and $\Delta \boldsymbol{\theta}_k = \boldsymbol{\theta}_k - \boldsymbol{\theta}_{k-1}$, the difference between two consecutive measurements. The elements of the $\mathbf{J}$ matrix by partial differentiation are:
\begin{subequations}
\label{eqn:elements_J}
\begin{align}
	\frac{\partial \text{P}_{m}}{\partial \theta_{n}} 
	& = \text{V}_{m} \text{V}_{n} \text{Y}_{m n} \sin \left( \theta_{m} - \theta_{n} - \alpha_{m n} \right) \,,  m \neq n \,,\label{eqn:elements_J_off}\\ 
	\frac{\partial \text{P}_{m}}{\partial \theta_{m}} 
	& = -\sum_{ n=1 \atop n \neq m}^{N} \frac{\partial \text{P}_{m}}{\partial \theta_{n}} \,. \label{eqn:elements_J_diag} 
\end{align}
\end{subequations}
Assuming that, under a normal operating condition, active power fluctuations are due to  random changes in electricity demand. We can model the active power mismatch by a Gaussian distribution, 
$
\Delta\textbf{P}_k \sim \mathcal{N}(\boldsymbol{0}, \sigma^2 \Delta t \mathbf{I}) \,,
$ where $\sigma^2$ is pre-determined and $\Delta t$ is the sampling interval. Therefore, we have
\begin{equation}
\label{eqn:angle_distribution}
    \Delta\boldsymbol{\theta}_k \sim  \mathcal{N}(\boldsymbol{0}, \sigma^2 (\mathbf{J}(\boldsymbol{\theta}_{k-1})^{T} \mathbf{J}(\boldsymbol{\theta}_{k-1}))^{-1}) \,.
\end{equation}

Suppose $K<N$ PMUs are installed on selected buses. Therefore, certain parts of the system are not directly observable, resulting in a degree of information loss as compared to the full PMU deployment case. In particular, for the $N - K$ buses, we do not observe their bus voltage phase angles and magnitudes. For a full PMU case, every element of the $\mathbf{J}$ matrix can be computed and updated with new PMU measurements. However, this would not be the case for a limited PMU deployment. For example, if there is no PMU installed on bus $m$, the off-diagonal element $\partial \text{P}_{m}/\partial \theta_{n}$ would not be computable, and the summation in the diagonal element $\partial \text{P}_{m}/\partial \theta_{m}$ is also affected. This inaccuracy in the $\mathbf{J}$ matrix has an impact on the effectiveness of the relationship (\ref{eqn:angle_distribution}) describing the system's dynamic behavior.

\subsection{Genetic Algorithm for Optimal Placement}
Let $S(n_p) = [x_1, \dots, x_N]$ denote a fixed PMU placement of $n_p$ PMUs on a power network of $N$ buses. In particular, $x_i = 0$ if the $i_{th}$ bus does not have a PMU, otherwise, $x_i = 1$, for $i = 1, \dots, N$. Given a fixed placement $S(n_p)$, we define the optimal placement to be the one that minimizes the discrepancy between the Jacobian matrix under a full PMU deployment, $\mathbf{J}(\boldsymbol{\theta})$, and the one under a limited PMU deployment, $\mathbf{J}_{S(n_p)}(\boldsymbol{\theta})$. Since the $\mathbf{J}$ matrix is time-variant and dependent on $\boldsymbol{\theta}$, we assume $\boldsymbol{\theta}$ follows a probability distribution $H$ on a close interval of $[-\pi, \pi]$. To quantify the discrepancy over the distribution of $\boldsymbol{\theta}$, we let 
\begin{equation}
\label{eqn:objective_function}
\delta_{S(n_p)} = \int_{\boldsymbol{\theta}} | \|\mathbf{J}(\boldsymbol{\theta})\|_{\mathrm{F}} - \|\mathbf{J}_{S(n_p)}(\boldsymbol{\theta})\|_{\mathrm{F}} | \, dH(\boldsymbol{\theta}) \,,
\end{equation} be the integral of the absolute difference between the Frobenius norms of the two matrices where the norm is defined as
\begin{equation}
\|\mathbf{A}\|_{\mathrm{F}} = \sqrt{\sum_{i=1}^{m} \sum_{j=1}^{n}\left|a_{i j}\right|^{2}} \,.
\end{equation}
Therefore, given a fixed number of PMUs, $n_p$, the OPP problem is now an optimization problem that can be written as
\begin{equation}
\begin{aligned}
\min_{S(n_p)} \quad	&	\delta_{S(n_p)}\\
\textrm{s.t.} \quad	&	S(n_p) = [x_1, \dots, x_N]\\
				&	x_i \in \{0,1\}\,, i = 1, \dots, N\\
				&	\sum_{i=1}^{N} x_i = n_p\\
\end{aligned}
\end{equation}

The actual distribution of the phase angles is unknown. One approach to circumvent this problem is to use the empirical distribution of $\boldsymbol{\theta}$ under the steady-state condition to approximate the otherwise intractable integration. To solve the combinatorial optimization problem, we use a meta-heuristic method, in particular, genetic algorithm (GA) to avoid the computational burden. GA is a type of optimization algorithms inspired by the natural reproduction and evolutionary process. These algorithms are more effective than random searches and more efficient than exhaustive searches \cite{kinnear1999advances}. The major components of a GA consist of a fitness function, an initial population, a mutation and crossover mechanism, and a next-generation selection mechanism. Using $\boldsymbol{\theta}$ and $\mathbf{V}$ under a normal condition, we evaluate the fitness of a given placement by (\ref{eqn:objective_function}). A mutation of one such individual is defined to be a position shuffle between a random pair of digits with a low probability. Finally, by repeatedly selecting the best one out of three random individuals from the current population, a fixed number of them forms the next generation. This step is known as the tournament. The process repeats for a fixed number of generations. In the last generation, the placement with the optimal fitness is chosen as the optimal solution. See Algorithm \ref{alg:GA} for the details of the GA procedure. 
\begin{algorithm}
\caption{Genetic algorithm for optimal placement}\label{alg:GA}
\begin{algorithmic}[1]
\Procedure{GA}{$\boldsymbol{\theta}$}
\State Generate $N_{ini}$ placements with $n_p$ PMUs\Comment{Create random initial population}
\State $g \gets 0$
\While{$g \not = N_{gen}$}\Comment{Run for $N_{gen}$ generations}
\State Evaluate $\delta_{S(n_p)}$ of every individual 
\State Select $N_{pop}$ individuals based on tournament of size 3
\ForAll{individuals}\Comment{Create next generation}
\State Mutate the individual with probability of $p_{mutate}$
\EndFor
\State $g \gets g + 1$
\EndWhile\label{alg:GAEndWhile}
\State \textbf{return} individual with the lowest $\delta_{S(n_p)}$\Comment{The optimal placement}
\EndProcedure
\end{algorithmic}
\end{algorithm}

\subsection{PMU Placement Evaluation}
Here we briefly describe the outage D\&I scheme developed in our previous work to evaluate the different PMU placements\footnote{For more details about the D\&I scheme, we refer the readers to \cite{yang2019control}.}. Since the $\mathbf{J}(\boldsymbol{\theta})$ matrix is determined by the network topology, a line outage would change the structure of the matrix. Let $\mathbf{J}_0(\boldsymbol{\theta})$ and $\mathbf{J}_{\ell}(\boldsymbol{\theta})$ represent the Jacobian with no outage and with outage at line $\ell$. We can formulate the detection problem as a sequential hypothesis testing problem with the following null and alternative hypothesis:
\begin{subequations}
\label{eqn:pre_post_distribution}
\begin{align}
 	 H_0: \Delta\boldsymbol{\theta}[k] &\sim \mathcal{N}(\boldsymbol{0}, \sigma^2 (\mathbf{J}_0(\boldsymbol{\theta})^T \mathbf{J}_0(\boldsymbol{\theta}))^{-1}) \,,  \label{eqn:pre_distribution}\\
 	 H_1: \Delta\boldsymbol{\theta}[k] &\sim \mathcal{N}(\boldsymbol{0}, \sigma^2 (\mathbf{J}_{\ell}(\boldsymbol{\theta})^T \mathbf{J}_{\ell}(\boldsymbol{\theta}))^{-1}) \,, \ell \in \mathcal{L} \,, \label{eqn:post_distribution}
\end{align}
 \end{subequations} where $\mathcal{L}$ is the set of all possible single-line outages. We adopt a generalized likelihood ratio (GLR) control chart approach, which repeatedly evaluates the likelihood of an outage against the likelihood of no outage. The GLR approach detects an outage at time $D$ where
\begin{equation}
\label{eqn:stopping_rule}
    D = \inf \left\lbrace  k \ge 1: \underset{\ell \in \mathcal{L}}{\max} \, W_{\ell, k} \ge c \right\rbrace \,.
\end{equation}
$W_{\ell, k}$ is the monitoring statistic of outage scenario $\ell$, and $c$ is a pre-determined threshold corresponding to a certain false alarm rate constraint. The threshold could be approximated by
\begin{equation}
\label{eqn:theshold}
c = \ln(ARL_0 \times p) \,,
\end{equation} where $ARL_0$ is the average run length to a false alarm under a normal operation and $p$ is the number of PMUs installed \cite{Chen2016}. Following the detection, we identify the tripped line by considering the top three probable candidates, $\ell_{(1)}, \ell_{(2)}, \text{and } \ell_{(3)}$ such that
\begin{equation}
\label{eqn:identification}
{W}_{\ell_{(1)}, D} \ge {W}_{\ell_{(2)}, D} \ge {W}_{\ell_{(3)}, D} \ge {W}_{y, D}\,,
\end{equation} for all $y \in \mathcal{L}$. 
For our placement evaluation, we are concerned with whether the outage can be detected, i.e., $D$ is less than the simulation duration, and whether the true tripped line is one of the three lines identified, i.e., $\ell \in \{\ell_{(1)}, \ell_{(2)}, \ell_{(3)} \}$.

\section{Simulation Studies}
\label{sec:simulation}
\subsection{Simulation Setting}
Assuming 20 PMUs are available, we test the GA-generated PMU placement as well as two other placements on single-line outages of the IEEE 39 bus New England system \cite{athay1979practical}. Outage dynamics are simulated using the open-source simulation platform COSMIC \cite{Song2016}. The sampling frequency of the installed PMUs is assumed to be 30 samples per second. For other outage-related simulation details, please refer to \cite{yang2019control}. For the GA, we set the number of generations to 50, the mutation probability of the individual to 0.2, and the bus index shuffling probability to 0.05. A tournament of size three is used to select the next generation population, set at a size of 100. Three other placement strategies are deployed, and their results are presented for comparison:
\begin{enumerate}
\item
Scattered placement: Assuming a random placement strategy, the PMUs are scattered across the whole power network.
\item
Tree placement: Assuming a connected placement strategy, the PMUs form a tree with no cycles in the context of a graph network. 
\item
Degree-based placement: The bus nodes are weighted in terms of network importance based on their degree of connection, e.g., a bus connected to six other buses has a degree of 6. Top 20 buses are equipped with a PMU. 
\end{enumerate}

\subsection{Simulation Results}
\begin{figure}[!h]
\centering
\includegraphics[width=1\linewidth]{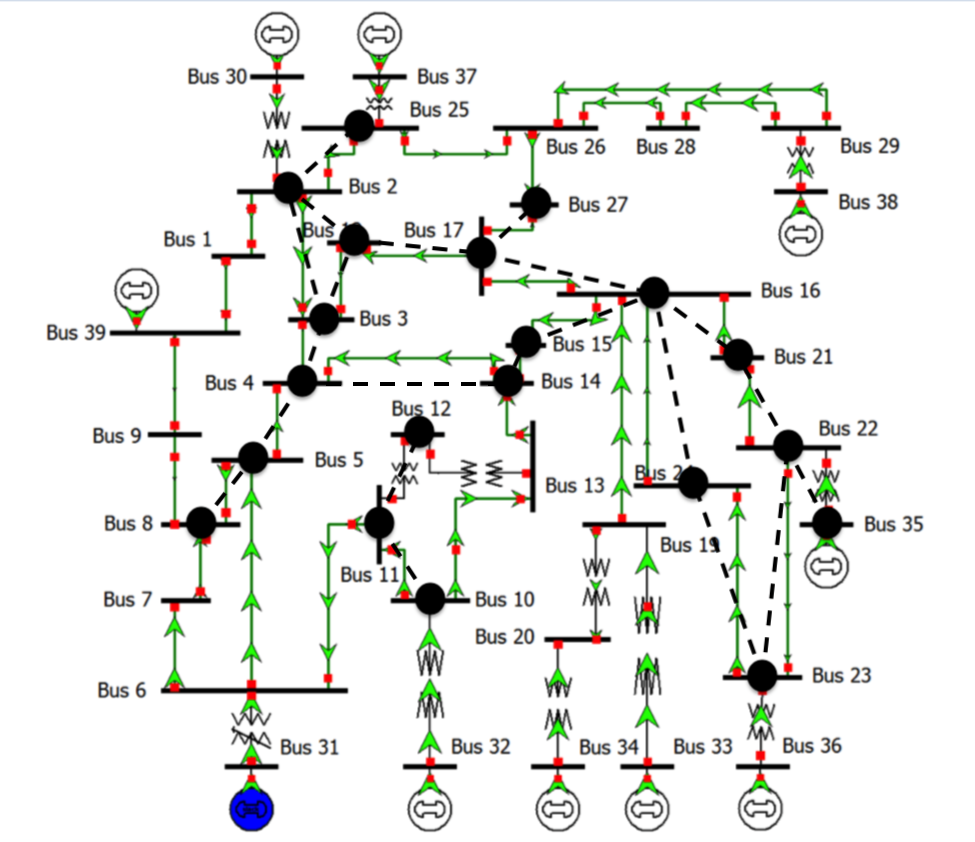}
\caption{Illustration of the GA-generated placement.}
\label{fig:IEEE39_genetic}
\end{figure}

See Table \ref{tab:placements} for details of the PMU placements under different strategies. Fig. \ref{fig:IEEE39_genetic} shows the GA-generated placement on the test system and it achieved an objective value where $\delta_{S(n_p)}^* = 147$. The GA program took, on average, 92 milliseconds with a standard deviation of 7.5 milliseconds to run on a desktop with a 2.9 GHz Intel Core i5 processor based on 1000 independent runs. See Fig. \ref{fig:IEEE39_scattered}, Fig. \ref{fig:IEEE39_tree}, and Fig. \ref{fig:IEEE39_degree} in the Appendix for illustration of other placement strategies\footnote{The topology plot is taken from the Illinois Center for a Smarter Electric Grid (ICSEG): https://icseg.iti.illinois.edu/ieee-39-bus-system/.}. The GA-generated placement resembles a spanning tree, albeit with cycles. Bus 16 and bus 18 have central positions in the graph, and they are connected through three and four edges, respectively. While the majority of the PMU buses are connected in a single graph, there is also a separate tree, i.e., bus 10, 11, and 12. 
\begin{table}[!t]
\renewcommand{\arraystretch}{1.3}
\caption{Placement Strategies and the Corresponding Placement}
\label{tab:placements}
\centering
\begin{tabular}{r|c}
\hline 
Placement Strategy & Placement (Bus) \\
\hline\hline
Scattered	& 1, 2, 5, 7, 9, 11, 13, 14, 16, 17, 19,\\& 21, 23, 24, 26, 27, 30, 32, 34, 37 \\
Tree		& 2-5, 7-9, 11-19, 21, 26-28 \\
Degree-based & 1-8, 10, 11, 13, 14, 16, 17, 19, 22, 23, 25, 26, 29 \\
GA-generated & 2-5, 8, 10-12, 14-18, 21-25, 27, 35\\
\hline
\end{tabular} 
\end{table} 

\subsubsection{Impact of the placement strategies}
\begin{figure}[!h]
\centering
\includegraphics[width=1\linewidth]{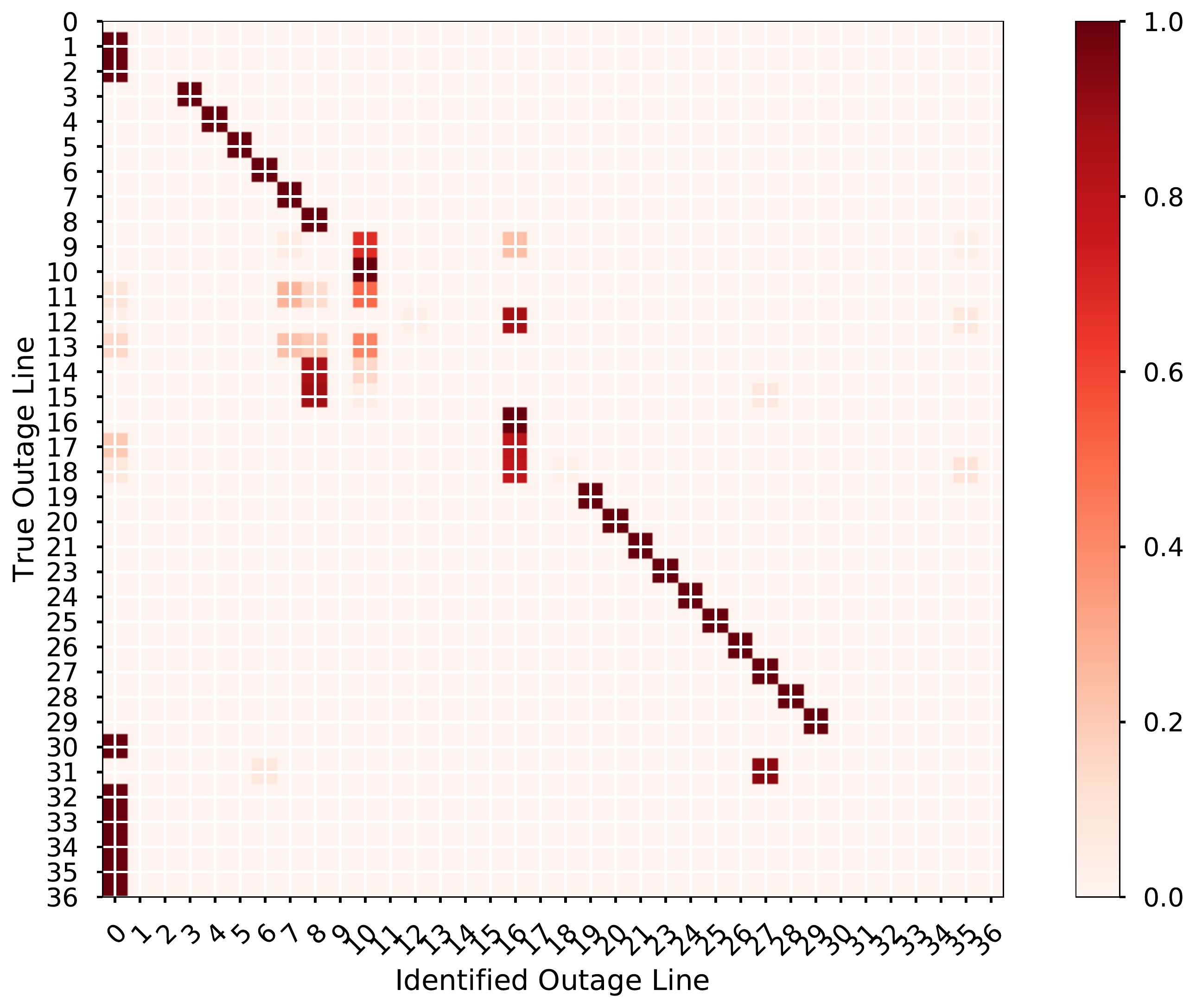}
\caption{Heat map showing the D\&I performance using the GA-generated placement strategy. The color bar on the right shows the percentage of the results based on 1000 Monte Carlo simulations; the darker the color, the higher the percentage. The horizontal axis shows the outage line number identified by the D\&I scheme whereas the vertical axis corresponds to the actual outage line.}
\label{fig:identification_top3_genetic_20pmu}
\end{figure}

We compare the outage D\&I performance of different PMU placements using a heat map with empirical likelihoods as entries. The horizontal axis represents the line identified by the identification scheme, and the vertical axis represents the actual tripped line. The cells of the heat map are color-coded based on the empirical likelihood of identifying the respective line outage based on 1000 line outage simulations. 0 on the horizontal axis indicates a missed detection. A perfect identification would have value 1 on all diagonal cells and 0 everywhere else. Fig. \ref{fig:identification_top3_genetic_20pmu} shows the performance of the placement found by the GA. The results for a full PMU deployment and a tree placement are shown in Fig. \ref{fig:identification_heatmap}, while the results for the other two strategies are omitted as they are significantly less effective. 

One aspect of the D\&I scheme performance is the likelihood of missed detection. It means how likely the detection scheme would miss the outage under a given PMU placement. They correspond to the first column of all the heat maps where actual outages are identified as $0$ by the detection scheme. The GA-generated placement shows a similar missed detection performance as compared to that of a full PMU deployment. There is a different degree of likelihood for missed detection across many lines under the tree placement, suggesting that a placement solely based on the topology graph connections might not be adequate for outage detection. The degree placement strategy results show similar evidence, which is not presented here.

Another aspect of the performance is the likelihood of correct identification. It means how likely the tripped line could be accurately identified. This aspect could be analyzed by looking at the diagonal entries of the heat maps. For the GA-generated placement, its performance matches that of the full deployment for most of the outages. Line 12 to 15 are not accurately identified, likely due to the lack of PMUs nearby. The scattered placement does not perform well, as many of the outages were not located. The reason is likely that many PMUs are isolated in this placement scenario, as seen in Fig. \ref{fig:IEEE39_scattered}. The isolation likely results in more inaccuracies between the Jacobian under a full and a partial PMU deployment. The tree placement, on the other hand, produced a decent identification performance with some inaccuracies towards the last part, as seen from Fig. \ref{fig:identification_top3_tree_20pmu}.

\subsubsection{Impact of the number of PMUs installed}
\begin{figure}[!ht]
\centering
\includegraphics[width=1\linewidth]{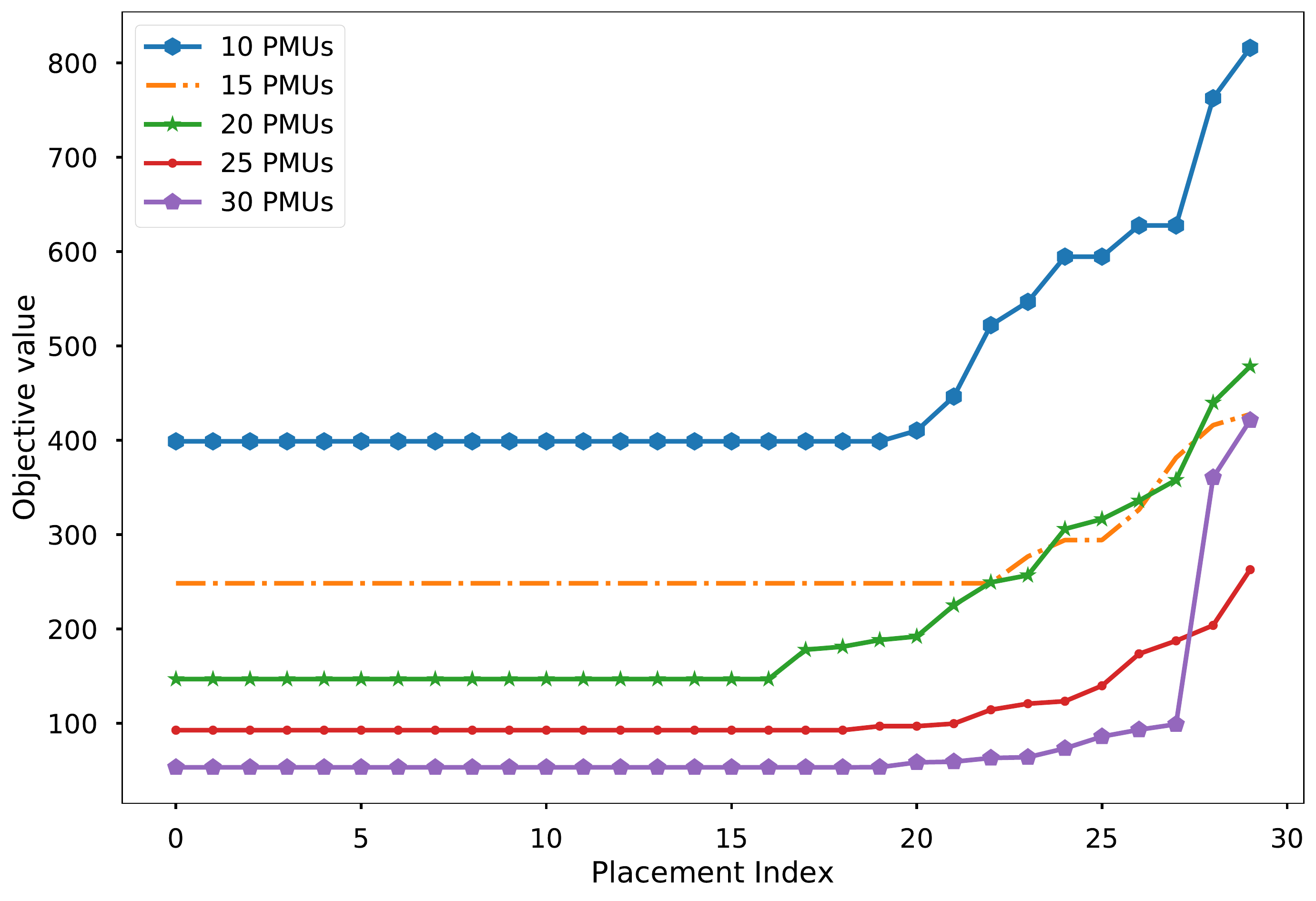}
\caption{Objective values of the top 30 placements in the last generation of the GA algotithm. Lines represent different number of PMUs installed. Objective values of the placements are sorted in an ascending order.}
\label{fig:obj_val_no_pmus}
\end{figure}

A fewer number of PMUs available corresponds to a higher degree of information loss. To quantify the impact, we implement the proposed GA for five different number of PMUs. Fig. \ref{fig:obj_val_no_pmus} shows the objective values of the best 30 placements found in GA respective to the given number of PMUs. Note that the individual placements in the last GA generation may not be unique, and many top placements are the same. Therefore, these placements give identical objective values, a feature obvious in Fig. \ref{fig:obj_val_no_pmus}. We can observe a considerable gap in objective values between the optimal placement and the non-optimal placements. The gap is especially significant when there are only 10 PMUs available. On the other hand, the objective values for optimal placements under the case of 20, 25, and 30 PMUs are close to each other, indicating a diminishing return to the number of PMUs. This phenomenon suggests that it would be worthwhile to investigate the minimum number of PMUs required to achieve a particular D\&I performance. 

\section{Conclusion}
\label{sec:conclusion}
In this work, we formulated the optimal PMU placement problem as an optimization problem that seeks to minimize the difference between the Jacobian matrix under a full PMU deployment case and a limited PMU deployment case. We adapted the GA to solve the optimization problem and illustrated the approach with the IEEE 39 bus system. The results show that the GA-generated placement has a better D\&I performance than the scattered, tree, and degree-based placement with 20 available PMUs. We also observed a diminishing return when more PMUs are available. The proposed method assumes a steady-state operating condition for the evaluation of the objective function. A further research direction could be incorporating power system dynamic models that describe the transient dynamics following an outage into the optimization problem.

\appendix
\section*{Comparison Placements and Results}

\begin{figure}[!h]
\centering
\includegraphics[width=1\linewidth]{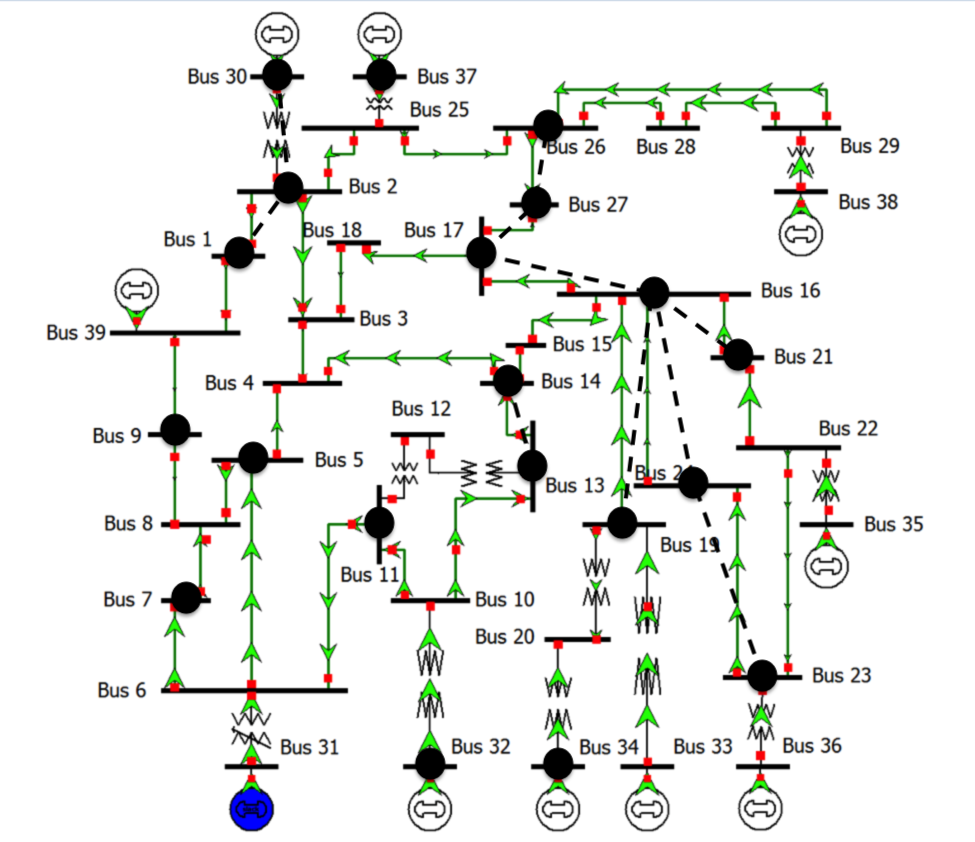}
\caption{Illustration of the scattered placement.}
\label{fig:IEEE39_scattered}
\end{figure}
\begin{figure}[!h]
\centering
\includegraphics[width=1\linewidth]{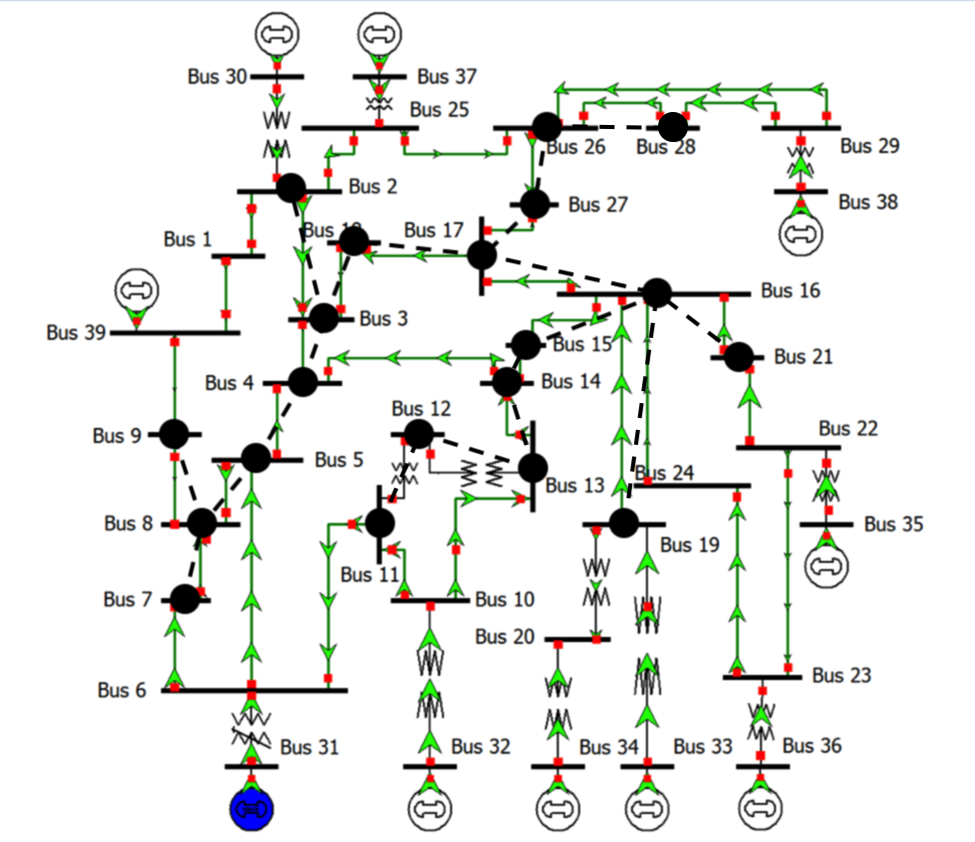}
\caption{Illustration of the tree placement.}
\label{fig:IEEE39_tree}
\end{figure}
\begin{figure}[!h]
\centering
\includegraphics[width=1\linewidth]{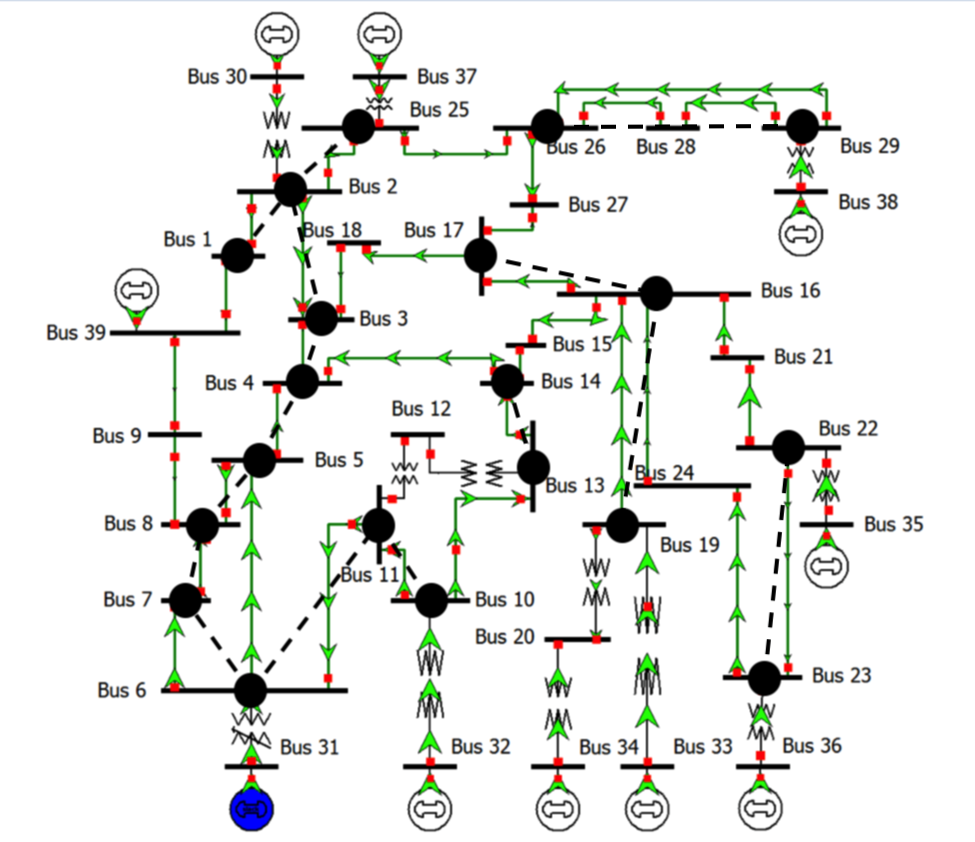}
\caption{Illustration of the degree-based placement.}
\label{fig:IEEE39_degree}
\end{figure}

\begin{figure}[!ht]
    \centering
  \subfloat[Full placement\label{fig:identification_full_top3}]{%
       \includegraphics[width=1\linewidth]{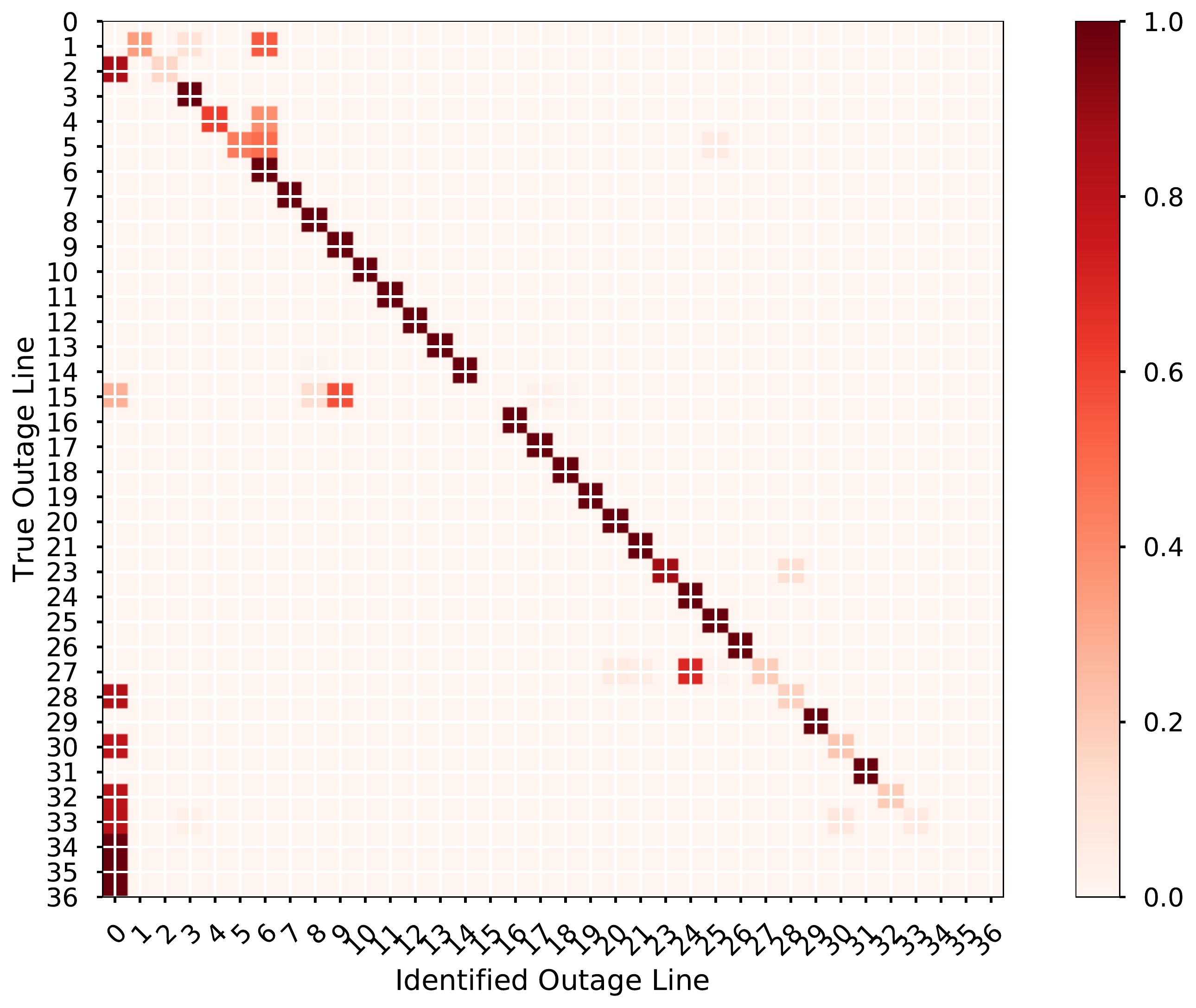}}\\
  \subfloat[Tree placement\label{fig:identification_top3_tree_20pmu}]{%
       \includegraphics[width=1\linewidth]{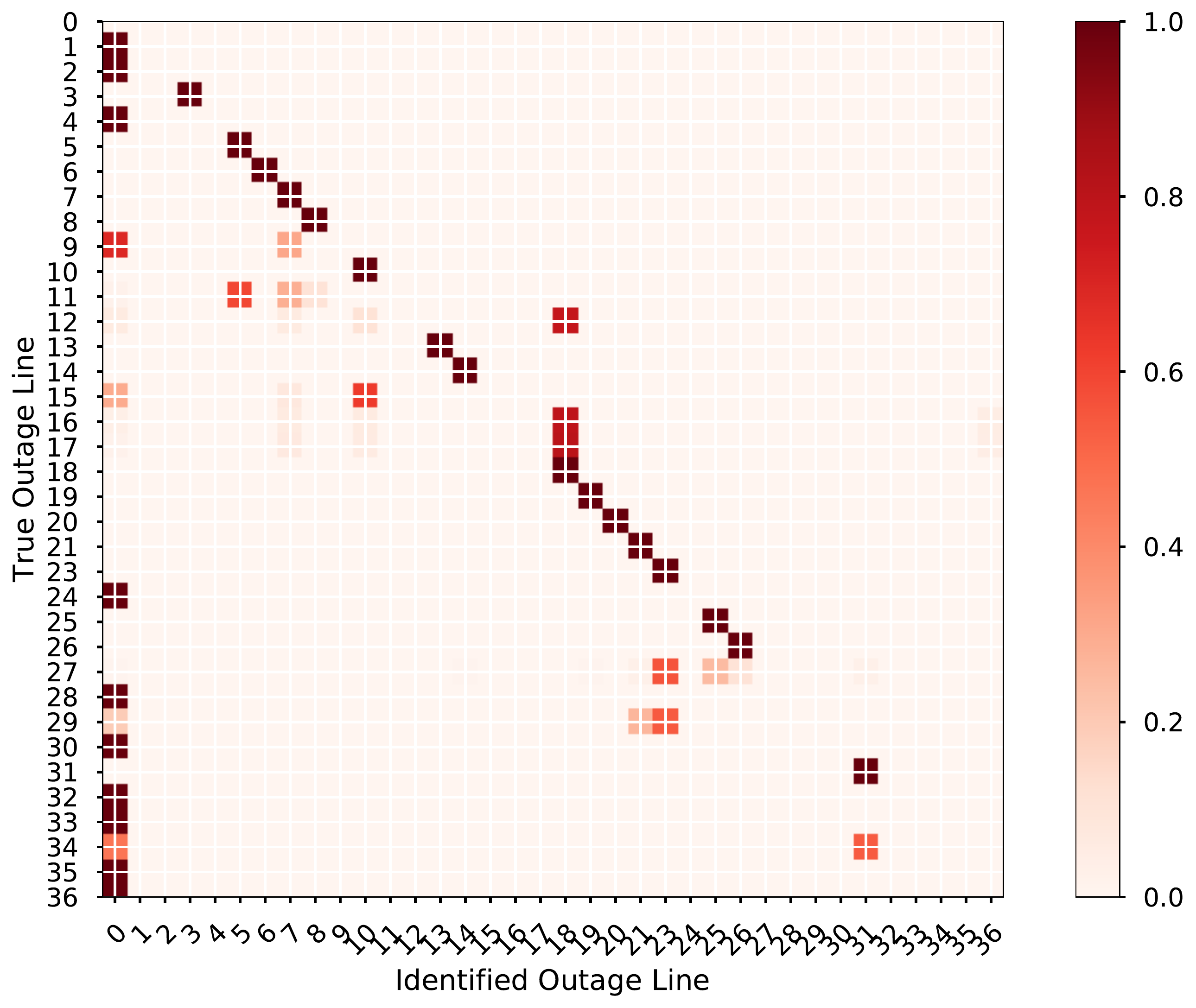}}
  \caption{Heat map showing the D\&I performance under full-PMU and tree placement strategies. The color bar on the right shows the percentage of the results based on 1000 Monte Carlo simulations; the darker the color, the higher the percentage. The horizontal axis shows the outage line number identified by the D\&I scheme, whereas the vertical axis corresponds to the actual outage line. a) Identification results for full placement. b) Identification results for tree placement.} 
  \label{fig:identification_heatmap}
\end{figure}


\begin{thebibliography}{10}
\providecommand{\url}[1]{#1}
\csname url@samestyle\endcsname
\providecommand{\newblock}{\relax}
\providecommand{\bibinfo}[2]{#2}
\providecommand{\BIBentrySTDinterwordspacing}{\spaceskip=0pt\relax}
\providecommand{\BIBentryALTinterwordstretchfactor}{4}
\providecommand{\BIBentryALTinterwordspacing}{\spaceskip=\fontdimen2\font plus
\BIBentryALTinterwordstretchfactor\fontdimen3\font minus
  \fontdimen4\font\relax}
\providecommand{\BIBforeignlanguage}[2]{{%
\expandafter\ifx\csname l@#1\endcsname\relax
\typeout{** WARNING: IEEEtran.bst: No hyphenation pattern has been}%
\typeout{** loaded for the language `#1'. Using the pattern for}%
\typeout{** the default language instead.}%
\else
\language=\csname l@#1\endcsname
\fi
#2}}
\providecommand{\BIBdecl}{\relax}
\BIBdecl

\bibitem{pignati2015real}
M.~Pignati, M.~Popovic, S.~Barreto, R.~Cherkaoui, G.~D. Flores, J.-Y.
  Le~Boudec, M.~Mohiuddin, M.~Paolone, P.~Romano, S.~Sarri \emph{et~al.},
  ``Real-time state estimation of the epfl-campus medium-voltage grid by using
  pmus,'' in \emph{2015 IEEE Power \& Energy Society Innovative Smart Grid
  Technologies Conference (ISGT)}.\hskip 1em plus 0.5em minus 0.4em\relax IEEE,
  2015, pp. 1--5.

\bibitem{Aminifar2014}
F.~Aminifar, M.~Fotuhi-Firuzabad, A.~Safdarian, A.~Davoudi, and
  M.~Shahidehpour, ``{Synchrophasor measurement technology in power systems:
  panorama and state-of-the-art},'' \emph{IEEE Access}, vol.~2, pp. 1607--1628,
  2014.

\bibitem{Tate2008}
J.~E. Tate and T.~J. Overbye, ``{Line outage detection using phasor angle
  measurements},'' \emph{IEEE Transactions on Power Systems}, vol.~23, no.~4,
  pp. 1644--1652, 2008.

\bibitem{Rafferty2016}
M.~Rafferty, X.~Liu, D.~M. Laverty, and S.~McLoone, ``{Real-time multiple event
  detection and classification using moving window PCA},'' \emph{IEEE
  Transactions on Smart Grid}, vol.~7, no.~5, pp. 2537--2548, 2016.

\bibitem{Chen2016}
Y.~C. Chen, T.~Banerjee, A.~D. Dominguez-Garcia, and V.~V. Veeravalli,
  ``{Quickest line outage detection and identification},'' \emph{IEEE
  Transactions on Power Systems}, vol.~31, no.~1, pp. 749--758, 2016.

\bibitem{Hosur2019}
S.~Hosur and D.~Duan, ``{Subspace-driven output-only based change-point
  detection in power systems},'' \emph{IEEE Transactions on Power Systems},
  vol.~34, no.~2, pp. 1068--1076, 2019.

\bibitem{Ardakanian2019a}
O.~Ardakanian, V.~W.~S. Wong, R.~Dobbe, S.~H. Low, A.~von Meier, C.~Tomlin, and
  Y.~Yuan, ``{on identification of distribution grids},'' \emph{IEEE
  Transactions on Control of Network Systems}, 2019.

\bibitem{yang2019control}
X.~Yang, N.~Chen, and C.~Zhai, ``A control chart approach to power system line
  outage detection under transient dynamics,'' \emph{arXiv:1911.01733
  [eess.SY]}, 2019.

\bibitem{marin2003genetic}
F.~Marin, F.~Garcia-Lagos, G.~Joya, and F.~Sandoval, ``Genetic algorithms for
  optimal placement of phasor measurement units in electrical networks,''
  \emph{Electronics Letters}, vol.~39, no.~19, pp. 1403--1405, 2003.

\bibitem{dua2008optimal}
D.~Dua, S.~Dambhare, R.~K. Gajbhiye, and S.~Soman, ``Optimal multistage
  scheduling of pmu placement: An ilp approach,'' \emph{IEEE Transactions on
  Power Delivery}, vol.~23, no.~4, pp. 1812--1820, 2008.

\bibitem{koutsoukis2013numerical}
N.~C. Koutsoukis, N.~M. Manousakis, P.~S. Georgilakis, and G.~N. Korres,
  ``Numerical observability method for optimal phasor measurement units
  placement using recursive tabu search method,'' \emph{IET Generation,
  Transmission \& Distribution}, vol.~7, no.~4, pp. 347--356, 2013.

\bibitem{Geramian2008}
S.~S. Geramian, H.~A. Abyane, and K.~Mazlumi, ``{Determination of optimal PMU
  placement for fault location using genetic algorithm},'' \emph{ICHQP 2008:
  13th International Conference on Harmonics and Quality of Power}, pp. 1--5,
  2008.

\bibitem{qi2014optimal}
J.~Qi, K.~Sun, and W.~Kang, ``Optimal pmu placement for power system dynamic
  state estimation by using empirical observability gramian,'' \emph{IEEE
  Transactions on power Systems}, vol.~30, no.~4, pp. 2041--2054, 2014.

\bibitem{gou2014unified}
B.~Gou and R.~G. Kavasseri, ``Unified pmu placement for observability and bad
  data detection in state estimation,'' \emph{IEEE Transactions on Power
  Systems}, vol.~29, no.~6, pp. 2573--2580, 2014.

\bibitem{rajeev2015fault}
A.~Rajeev, T.~Angel, and F.~Z. Khan, ``Fault location in distribution feeders
  with optimally placed pmu's,'' in \emph{2015 International Conference on
  Technological Advancements in Power and Energy (TAP Energy)}.\hskip 1em plus
  0.5em minus 0.4em\relax IEEE, 2015, pp. 438--442.

\bibitem{Jamei2017a}
M.~Jamei, A.~Scaglione, C.~Roberts, E.~Stewart, S.~Peisert, C.~McParland, and
  A.~McEachern, ``{Anomaly detection using optimally placed µPMU sensors in
  distribution grids},'' \emph{IEEE Transactions on Power Systems}, vol.~33,
  no.~4, pp. 3611--3623, 2017.

\bibitem{Glover2012}
J.~D. Glover, M.~S. Sarma, and T.~Overbye, \emph{{Power System Analysis {\&}
  Design, SI Version}}.\hskip 1em plus 0.5em minus 0.4em\relax Cengage
  Learning, 2012.

\bibitem{kinnear1999advances}
K.~E. Kinnear, W.~B. Langdon, L.~Spector, P.~J. Angeline, and U.-M. O'Reilly,
  \emph{Advances in genetic programming}.\hskip 1em plus 0.5em minus
  0.4em\relax MIT press, 1999, vol.~3.

\bibitem{athay1979practical}
T.~Athay, R.~Podmore, and S.~Virmani, ``A practical method for the direct
  analysis of transient stability,'' \emph{IEEE Transactions on Power Apparatus
  and Systems}, no.~2, pp. 573--584, 1979.

\bibitem{Song2016}
J.~Song, E.~Cotilla-Sanchez, G.~Ghanavati, and P.~D.~H. Hines, ``{Dynamic
  modeling of cascading failure in power systems},'' \emph{IEEE Transactions on
  Power Systems}, vol.~31, no.~3, pp. 2085--2095, 2016.

\end{thebibliography}
\end{document}